\documentclass[conference]{IEEEtran}

  \usepackage{caption}
  \usepackage{comment}
  \usepackage{lipsum}
\usepackage{cite}
\usepackage{amsmath,amssymb,amsfonts}
\usepackage{algorithmic}
\usepackage[applemac]{inputenc} 
\usepackage{graphicx}
\newcommand{\RN}[1]{%
  \textup{\uppercase\expandafter{\romannumeral#1}}%
}
\newcommand*\sepline{%
   \begin{center}
     \rule[1ex]{0.9\textwidth}{.4pt}
   \end{center}}
\usepackage{float}
\usepackage{cuted,nccmath}
\usepackage{flushend}
 \usepackage{booktabs}
 \usepackage{multirow}
 \usepackage[table,xcdraw]{xcolor}
\usepackage{textcomp}
\usepackage{mathtools, nccmath}
\usepackage{xcolor}
\def\BibTeX{{\rm B\kern-.05em{\sc i\kern-.025em b}\kern-.08em
    T\kern-.1667em\lower.7ex\hbox{E}\kern-.125emX}}
\begin{document}
\title{PLS for V2I Communications Using Friendly Jammer and Double kappa-mu Shadowed Fading
{\footnotesize \textsuperscript{ }}
\thanks{ }
}

\author{\IEEEauthorblockN{\textsuperscript{} Neji Mensi and Danda B. Rawat}
\IEEEauthorblockA{\textit{Department of Electrical Engineering and Computer Science} \\
\textit{Howard University}\\
Washington, DC 20059, USA \\
neji.mensi@bison.howard.edu, danda.rawat@howard.edu }
\and
\IEEEauthorblockN{\textsuperscript{} Elyes Balti}
\IEEEauthorblockA{\textit{Wireless Networking and Communications Group} \\
\textit{The University of Texas at Austin}\\
Austin, TX 78712, USA\\
ebalti@utexas.edu}

}

\maketitle

\begin{abstract}
The concept of intelligent transportation systems (ITS) is considered to be a highly promising area of research due to its diversity of unique features. It is based mainly on the wireless vehicular network (WVN), where vehicles can perform sophisticated services such as sharing real-time safety information. To ensure high-quality service, WVN needs to solve the security challenges like eavesdropping, where malicious entities try to intercept the confidential transmitted signal. In this paper, we are going to provide a security scheme under the Double kappa-mu Shadowed fading. Our solution is based on the use of a friendly jammer that will transmit an artificial noise (AN) to jam the attacker's link and decrease its eavesdropping performances.  To evaluate the efficiency of our solution, we investigated the outage probability for two special cases: Nakagami-m and Rician shadowed while taking into consideration the density of the blockage and the shadowing effects. We also studied the average secrecy capacity via deriving closed-form expressions of the ergodic capacity at the legitimate receiver and the attacker for the special case: Nakagami-m fading distribution.

\end{abstract}

\begin{IEEEkeywords}
Physical Layer Security, Eavesdropping, Wireless Vehicular Network, Artificial Noise, Double $\kappa$-$\mu$ Shadowed Fading.
\end{IEEEkeywords}

\section{Introduction}
\subsection{Background and Literature Review}
Wireless vehicular network (WVN) has been exponentially evolving by taking advantage of sophisticated technologies such as artificial intelligence (AI) \cite{ai}, machine learning \cite{ml}, Millimeter Wave (mmWave) \cite{eJEns_Mmwave,eRelayFSO3, mmw,eAppr,mmwArx }, and 5G \cite{5g,eRelayFSO2,eRelayFSO1}, etc.  It offers a variety of advanced features like real-time alerting messages \cite{realTime}, cloud services \cite{cloud, eVCloud}, etc.  However, as any communication network, WVN  is subject to many challenges, especially the security issue which is a peer parameter that guarantees a certain acceptable quality of services (QoS). \\
Several research papers discussed the security challenges and proposed a variety of solutions from different OSI (Open Systems Interconnection model) layers perspective \cite{generalSec1,generalSec2}. Regarding the physical layer security (PLS), it has been proven that securing the communication at this level is an efficient scheme to deal with passive threats like the eavesdropping attacks. In this kind of menace, the malicious network entity is able to listen to a private communication by intercepting the transmitted signal and revealing the confidential information.\let\thefootnote\relax\footnotetext{  This work was supported in part by the U.S. National Science Foundation (NSF) under the grant CNS-1650831.} What makes the eavesdropping attacks very critical is the fact that it is difficult to be detected by the victims since it could be processed without leaving any traces.
To deal with this security dilemma,  some related works proposed the use of artificial noise (AN) where the transmitter perturbs the attacker channel by sending a dedicated signal \cite{an1,an2}. This signal is generated orthogonally to the main link between the legitimate entities so that it will affect only the eavesdropper link. Other papers studied the employment of a friendly jammer (J) which has the responsibility of jamming the eavesdroppers' channels instead of the transmitter \cite{jam1,jam2}. \\
In general, by employing a jammer $J$, as a third network entity, could be more efficient in the case of multiple eavesdroppers attacks. In other words, if the network is subject to several attacks, it is better to have a jammer node to deal with all the attackers rather than each communicating node deals with all the attackers by itself (it will be very expensive in terms of power if each transmitter will dedicate a fraction of its power to jam all the eavesdroppers' channels). 
\subsection{Our Contribution}
In our paper, we are focusing on using the friendly jammer to protect V2I communications. As a channel fading, we adopt the Double Shadowed $\kappa$-$\mu$ Fading Model, noted by $\mathcal{D}(\cdot)$, recently presented in \cite{channel}, which is more general and covers wide ranges of fading such as double shadowed Rice, Rician shadowed, Nakagami-q, Nakagami-m, Rayleigh, one-sided Gaussian, etc. 
 We highlight our contributions as follows:
\begin{enumerate}
  \item[$\bullet$] We propose the use of friendly jammer $J$  in V2I communications under an eavesdropping attack scenario.
   \item[$\bullet$] We adopt the new channel model $\mathcal{D}(\cdot)$.
    \item[$\bullet$] We derive a closed form expression for the ergodic capacity at the receiver under $\mathcal{D}(\cdot)$ model.
  \item[$\bullet$] We derive a closed form expression for cumulative distribution function (CDF) of the signal-to-interference-plus-noise-ratio (SINR) and the ergodic capacity at the eavesdropper while considering Nakagami-m as special case of $\mathcal{D}(\cdot)$.
  \item[$\bullet$] We examine the impact of the blockage density at the receiver by adopting the special case model Rician shadowed.
\end{enumerate}
\subsection{Paper Structure}
The paper is constructed as follows:  Section \RN{2} introduces the system model. Section \RN{3} studies the outage probability while Section \RN{4}  examines the ergodic capacity and the secrecy capacity. Then, Section \RN{5} evaluates the performance of the security approach based on numerical results. Finally, we outline our conclusion in Section \RN{6}.
\section{System Model}
\subsection{Vehicular Communications and Attack Model}
\begin{figure}[H]
\includegraphics[height=55mm, width=\linewidth]{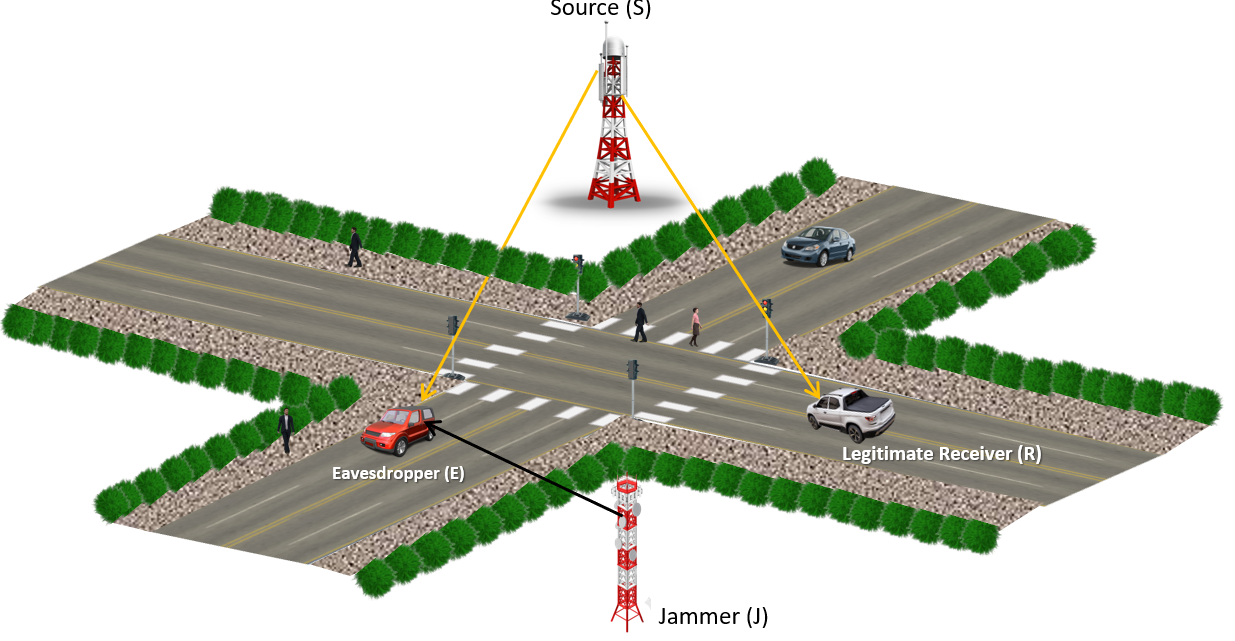}
\caption{V2I communications in the presence of multiple eavesdroppers.  }
\end{figure}
In V2I communications, base stations (BSs) and vehicles should be able to exchange information and data securely. However, the transmitted signals may be subject to an intended overhearing where an eavesdropper intercepts the signal and reveals the secret messages. In Fig. 1, we have a typical model where the attacker \textit{E} is listening to the communication between the source $S$ (the base station) and the legitimate receiver $R$. In this case,  $J$  tends to protect the network by sending AN to the attacker $E$, while $R$ is immune since the AN is orthogonal to its channel.
\subsection{Channel Model}
The received signal at \textit{R} and \textit{E}  are respectively:
\begin{equation}
\begin{aligned}
{ y_{_R}}& = \sum_{n=1}^{N}h_{_{S,R;n}}x_{_I} + \sum_{k=1}^{K}h_{_{J,R;k}}x_{_J} +w_{_R}\\
&= \sum_{n=1}^{N}h_{_{S,R;n}}x_{_I} +w_{_R},
\end{aligned}
\end{equation}
\begin{equation}
\begin{aligned}
{ y_{_E}}& =\sum_{n=1}^{N}h_{_{S,E;n}}x_{_I} + \sum_{k=1}^{K}h_{_{J,E;k}}x_{_J} +w_{_E},  
\end{aligned}
\end{equation}
where the channel model parameters are defined in Table I.
\begin{table}[h]
\caption{ Channel model parameters description} 
\begin{tabular}{|c|c|}
\hline
\rowcolor[HTML]{FFCE93} 
Parameters                                                                     & \multicolumn{1}{c|}{\cellcolor[HTML]{FFCE93}Description}                                                                                                                                                 \\ \hline
N                                                                              & \multicolumn{1}{c|}{Number of antennas at the BS}                                                                                                                                                         \\ \hline
\rowcolor[HTML]{EFEFEF} 
K                                                                              & \multicolumn{1}{c|}{\cellcolor[HTML]{EFEFEF}Number of antennas at the Jammer \textit{J }}                                                                                                                           \\ \hline
                                                                   $h_{_{a,b;c}}$            & \multicolumn{1}{c|}{\begin{tabular}[c]{@{}c@{}}The fading amplitude of the channel corresponding\\ to the link  between the antenna $c$ of the node $a$\\ and the receiving node $b$.\\ $a\in\{J,S\}$ and $b\in\{R,E\}$\end{tabular}} \\\hline \cline{2-2} 
                                                                               & \multicolumn{1}{c|}{\cellcolor[HTML]{EFEFEF}$g_{_{a,b}}$ is the channel gain where  $|g_{a,b}|^2$ $\sim \mathcal{D}(\cdot)$}                                                                                                                                                \\ \cline{2-2} 
                                                                               & \multicolumn{1}{c|}{\begin{tabular}[]{@{}c@{}}$r_{_{a,b;c}}$ is the distance between the antenna $c$ \\of the node $a$ and the receiving node $b$\end{tabular}}                                                                                                                                                                                   \\ \cline{2-2} 
\multirow{-4}{*}{\begin{tabular}[c]{@{}c@{}}$h_{_{a,b;c}}$\\  $=g_{_{a,b;c}}\sqrt{r_{_{a,b;c}}^{-\delta}}$\end{tabular}} & \cellcolor[HTML]{EFEFEF}$\delta$ is path loss exponent                                                                                                                                                                        \\ \hline
\rowcolor[HTML]{EFEFEF} 
$x_{_I}$                                                                           & \multicolumn{1}{c|}{\cellcolor[HTML]{EFEFEF}\begin{tabular}[]{@{}c@{}}The confidential information signal\\ sent by $S$ with power $P_{_S}$/antenna \end{tabular}}                                                                                                                                                            \\ \hline
$x_{_J}$                                                                           & \multicolumn{1}{c|}{\begin{tabular}[]{@{}c@{}}jamming signal (AN) emitted by $J$\\ with power $P_{_J}$/antenna \end{tabular}}    
\\ \hline
\rowcolor[HTML]{EFEFEF} 
$w_{_b}$                                                                          & \multicolumn{1}{c|}{\cellcolor[HTML]{EFEFEF}\begin{tabular}[]{@{}c@{}}The additive white Gaussian noise (AWGN) \\at the node $b$ with variance ${\sigma_{w_{_b}}^2}$ \end{tabular}}                                                                                                                                                       \\ \hline
\end{tabular}
\end{table}

As we can deduce from Eq. (1), $\sum_{k=1}^{K}h_{_{J,R}}x_{_J}=0$, which means that the jamming signal will only affect the attacker while conserving the same signal at the legitimate receiver.\\
The received SNR and SINR at $R$ and $E$, respectively, are
\begin{equation}
\begin{aligned}
{ \gamma_{_R}}=\frac{ \sum_{n=1}^{N}P_{_S}|g_{_{S,R}}|^{2}r_{_{S,R;n}}^{-\delta} }{\sigma_{w_{_R}}^2} = \sum_{n=1}^{N}\gamma_{_{R;n}}, 
\end{aligned}
\end{equation}
\begin{equation}
\begin{aligned}
{ \gamma_{_E}} =&\frac{\sum_{n=1}^{N}P_{_S}|h_{_{S,E;n}}|^{2} }{\sum_{k=1}^{K}P_{_{J}}|h_{_{J,E;k}}|^{2}+{\sigma_{w_{_E}}^2}}=\frac{ \frac{\sum_{n=1}^{N}P_{_S}|g_{_{S,E;n}}|^{2}r_{_{S,E;k}}^{-\delta}}{\sigma_{w_{_E}}^2}} {1+\frac{\sum_{k=1}^{K}P_{_{J}}|g_{_{J,E;k}}|^{2}r_{_{J,E;k}}^{-\delta}}{{\sigma_{w_{_E}}^2}}}\\&=\frac{ \sum_{n=1}^{N}\gamma{_{_{I;n}}} }{1+\sum_{k=1}^{K}\gamma{_{_{J;k}}}}=\frac{ \gamma{_{_{I}}} }{1+\gamma{_{_{J}}}}, 
\end{aligned}
\end{equation}
where $\gamma_{_{R;n}} $ $\sim$ $\mathcal{D}(\cdot)$ and $\gamma{_{_{I;n}}}$ $\sim$ $\mathcal{D}(\cdot)$ are, respectively,  the  SNRs  corresponding  to  the  confidential signal received at $R$ and $E$ via the $n$-th antenna, while $\gamma{_{_{J;k}}}$ $\sim$ $\mathcal{D}(\cdot)$ is the SNR due to the jamming signal sent by $J$ at $E$ via the $k$-th antenna.
The general probability density function (PDF) and CDF of the random variable (RV) $\gamma_{_l}$, where $ l\in \{(I;n), (J;k), (R;n) \}$, are respectively
\begin{equation}
\begin{aligned}
{ f}_{\gamma_{_l}}(\gamma) &=\frac{(s_{_l}-1)^{s_{_l}}c_{_l}^{c_{_l}}T_{_l}^{\mu_{_l}}\gamma^{\mu_{_l}-1}\overline{\gamma}^{s_{_l}}}{(c_{_l}+\mu_{_l}\kappa_{_l})^{c_{_l}}B(s_{_l},\mu_{_l})(T_{_l}\gamma+(s_{_l}-1)
\overline{\gamma})^{s_{_l}+\mu_{_l}}}\\
&\times_{2}F_{1}\left( c_{_l}, s_{_l}+\mu_{_l};\mu_{_l}; \frac{K_{_l}\mu_{_l}\kappa_{_l}\gamma}{T_{_l}\gamma+(s_{_l}-1)\overline{\gamma}}\right),
\end{aligned}
\end{equation}
\begin{equation}
\begin{aligned}
&{ F}_{\gamma_{_l}}(\gamma) =\left( \frac{c_{_l}}{c_{_l}+\kappa_{_l}\mu_{_l}}\right)^{c_{_l}}
\left( \frac{T_{_l}\gamma}{\overline{\gamma}(s_{_l}-1)}\right)^{\mu_{_l}} \sum_{i=0}^{\infty} \left(\frac{K_{_l}\mu_{_l}\kappa{_l}\gamma}{(s_{_l}-1)\overline{\gamma}}\right)^i
\\& \frac{(c_{_l})_i+(i+\mu_{_l})_{s_{_l}}}{i!\Gamma(s_{_l})(i+\mu_{_l})}
{}_{2}F_{1}( i+\mu_{_l},i+\mu_{_l}+ s_{_l};i+\mu_{_l}+1;\mu_{_l}; \tau),
\end{aligned}
\end{equation}
where $_{2}F_{1}(\cdot,\cdot;\cdot;.)$ represents the Hypergeometric function, $B(\cdot,\cdot)$ denotes the Beta function, $\tau=\frac{-T_{_l}\gamma}{\overline{\gamma}(s_{_l}-1)}$ and $(x)_{i}=\frac{\Gamma(x+i)}{\Gamma(x)}$ is the  the Pochhammer symbol. For the sake of organization, the parameters on which depends the distribution of $\gamma_{_l}$, are presented in Table. II.
\begin{table}[h]
\caption{ $\mathcal{D}(\cdot)$ parameters description\cite{channel}} 
\begin{tabular}{|r|c|c|}
\hline
\rowcolor[HTML]{FFCE93} 
\multicolumn{1}{|c|}{\cellcolor[HTML]{FFCE93}SNR} & Parameters                    & Description                                                    \\ \hline
                                                  & $c_{_l}$                            & Shape of the Nakagami-m RV                                     \\ \cline{2-3} 
                                                  & \cellcolor[HTML]{EFEFEF}$s_{_l}$    & \cellcolor[HTML]{EFEFEF}Shape of  the inverse of Nakagami-m RV \\ \cline{2-3} 
                                                  & $\mu_{_l}$                            & Number of multipath clusters                                   \\ \cline{2-3} 
                                                  & \cellcolor[HTML]{EFEFEF}$\kappa_{_l}$ & \cellcolor[HTML]{EFEFEF}\begin{tabular}[]{@{}c@{}}The ratio of the total power of the dominant\\
components to the scattered waves\end{tabular}                       \\ \cline{2-3} 
                                                  & $T_{_l}$                             &   $\mu_{_l}(1+\kappa_{_l})$                                                             \\ \cline{2-3} 
\multirow{-6}{*}{\begin{tabular}[c]{@{}r@{}}$\gamma_{_l} $\end{tabular}}           & \cellcolor[HTML]{EFEFEF}$K_{_l}$    & \cellcolor[HTML]{EFEFEF} $\frac{K_{_l}}{(c_{_l}+\mu_{_l}*\kappa_{_l})}$                                      \\ \hline
\end{tabular}
\end{table}

\section{Outage Probability Analysis}
\subsection{Outage Probability at the Legitimate Receiver }
To analyze the outage probability at $R$, we have to refer to the corresponding CDF. We know that  $\gamma_{_{R;n}} $ $\sim$ $\mathcal{D}(\cdot)$, hence its CDF is given by Eq. (6). However, the closed-form expression of the CDF corresponding of $\sum_{n=1}^{N}\gamma_{_{R;n}} $ is not tractable. Therefore, we are assuming that N = 1, which make the outage probability of $\gamma_{_{R}}$ represented by Eq. (6). At this stage, we are referring to the  special fading case Rician shadowed under the following substitutions: $c_{R}\rightarrow\infty$ and $\mu = 1$  \cite{channel}. Therefore, the CDF is expressed as follows \cite{ShadRicianCDF}
\begin{equation}
\begin{aligned}
{ F}_{\gamma_{_R}}(\gamma) = \frac{1}{\Gamma(m)}\left(\frac{m}{\Xi}\right)^m \sum_{i=0}^{\infty}\frac{\Gamma(m+i)\gamma\left(i+1,\frac{\gamma}{\overline{\gamma}2\sigma^2}\right)}{\sigma^{2i}2^{i}i!\Gamma(1+i)\left(\frac{1}{2\sigma^{2}}+\frac{m}{\Xi}\right)^{m+i}},
\end{aligned}
\end{equation}
where $\gamma(\cdot,\cdot)$ is the Incomplete Gamma function, $\Xi$ is the average power of the line of sight (LOS) component, $2\sigma^2$ is the average power of the scatter
component, $m$ is  fading figure which represents the fading severity.
Therefore, the outage related to the density of shadowing, $F_{\gamma_{_{SD}}}$ could be presented by
\begin{equation}
\begin{aligned}
 F_{\gamma_{_{SD}}}(\gamma) = p_{_{los}}F_{\gamma_{_{SD}}}^{_{los}}(\gamma) +(1-p_{_{los}})F_{\gamma_{_{SD}}}^{_{nlos}}(\gamma) ,
\end{aligned}
\end{equation}
where $F_{\gamma_{_{SD}}}^{_{los}}(\gamma)$ and $F_{\gamma_{_{SD}}}^{_{nlos}}(\gamma)$ are the CDFs of the  SNR evaluated when the link is LOS, and non-line-of-side (NLOS), respectively.
\subsection{Outage Probability at the Attacker }
It is complex to derive a closed-form expression of the CDF at the eavesdropper $E$. However, we can take advantage of the $\mathcal{D}(\cdot)$ generality by considering special channel model cases by manipulating its parameters. 
 We suggest that $h_{S,E}$ and $h_{J,E}$ follow  Nakagami-m distribution  by fixing $c_{_R}=s_{_R}=\infty$, $\mu_{_I}=1$ and $\kappa_{_I}=m$\cite{channel}. Hence,  $\gamma_{_{I;n}}$ and $\gamma_{_{J;k}}$ (see Eq. (4) )  are described by Gamma distribution ($\gamma_{_{I;n}} \sim\textit{\textsf{G}}(\nu_{_{I,n}},\beta_{_{I,n}})$ and $\gamma_{_{J;k}} \sim\textit{\textsf{G}}(\nu_{_{J,k}},\beta_{_{J,k}})$ ) , which has the following PDF and CDF
\begin{equation}
\begin{aligned}
{ f}_{\gamma_{_d}}(\gamma) =\frac{\beta_{_d}^{\nu_{_d}}\gamma^{\left(\nu_{_d}-1\right)}\exp({-\beta_{_d}\gamma})}{\Gamma(\nu_{_d})},
\end{aligned}
\end{equation}
\begin{equation}
\begin{aligned}
{ F}_{\gamma_{_d}}(\gamma) &  =1-\frac{\Gamma(\nu_{_d},\beta_{_d}\gamma) }{\Gamma(\nu_{_d})}, 
\end{aligned}
\end{equation}
where $d\in\{(I;n),(J;k)\}$, $\beta_{_h}$ and $\nu{_h}$ are, respectively, the scale and the shape parameters. Each antenna of the BS  has the same distance from the attacker because the antennas are collocated. Therefore the parameter $\beta_{_{I,n}}$ is the same for all antennas $n\in {1,...,N}$ ($\beta{_{I}}$=$\beta_{_{I;n}}$) since $\beta_{_{S,E;n}}=\frac{ P_{_S} r_{_{S,E;n}}^{-\delta}}{\sigma_{w_{_E}}^2}$. We assume that all \textit{N} links have the same scale parameter, which means $\nu{_{I}}$=$\sum_{n=1}^{N}\nu_{_{I;n}}$.  Therefore $\sum_{n=1}^{N} \gamma_{_{I;n}} = \gamma_{_{I}}$ $\sim$ $\textit{\textsf{G}}(\nu_{_{I}},\beta_{_{I}})$. The same analogy applied to the signal issued by the jammer where $\sum_{k=1}^{K} \gamma_{_{J;k}} = \gamma_{_{J}}$ $\sim$ $\textit{\textsf{G}}(\nu_{_{J}},\beta_{_{J}})$.
For the sake of mathematical simplicity, we assume that $\nu_{_h}$ is a positive integer. Accordingly, the CDF has the subsequent series expansion as:
\begin{equation}
\begin{aligned}
{ F}_{\gamma_{_d}}(\gamma) =1- \sum_{n=0}^{\nu_d-1} \frac{(\beta_{_d}\gamma)^{n} e^{-\beta_{_d}\gamma}}{n!} .
\end{aligned}
\end{equation}
Before deriving the CDF at $E$ following the aforementioned special cases, we should mention that we verified and proved obtaining  Nakagami-m distribution by  using the general  PDF of the envelope corresponding to $\mathcal{D}(\cdot)$ and substituting the appropriate parameters (Please refer to the Appendix).\\
By referring to Eq. (4), Eq. (9), and Eq. (11), the CDF at the attacker can be derived using the following expression

\begin{equation}
\begin{aligned}
&F_{\gamma_{E}}=\int\limits_{0}^{\infty}F_{\gamma_{_I}}(\gamma[1+\gamma_{_J}])f_{\gamma_{_J}}(\gamma_{_J})d\gamma_{_J}\\
&=1-\frac{e^{-\beta_{_I}\gamma}{\beta_{_J}}^{\nu_{_J}}}  {\Gamma(\nu_{_J})}\sum_{n=0}^{\nu_{_I}-1}\sum_{q=0}^{i}{n\choose q}\int\limits_{0}^{\infty}\frac{\gamma_{_J}^{\Omega-1} e^{-\gamma_{_J}(\beta_{_I} \gamma+\beta_{_J})}}{n!(\beta_I\gamma)^{-n} }d\gamma_{_J}
\end{aligned}
\end{equation}
where $\Omega=q+\nu_{_J}$.\\
Then, by referring to [\citenum{Tab}, Eq (3.351.3)], we obtain
\begin{equation} 
\begin{aligned}
{ F}_{\gamma_{_E}}(\gamma) &=1-\frac{e^{-\beta_{_I}\gamma}{\beta_{_J}}^{\nu_{_J}}}  {\Gamma(\nu_{_J})}\sum_{n=0}^{\nu_{_I}-1}\sum_{q=0}^{i}{n\choose q}\frac{\Gamma(\Omega)(\beta_{_I}\gamma+\beta_{_J})^{-\Omega}}{n!(\beta_I\gamma)^{-n}}
\end{aligned}
\end{equation}

\section{Average Secrecy Capacity  Analysis}
The generalized formula of the ergodic capacity for a given SNR $\gamma$ is expressed by
\begin{equation}
\begin{aligned}
   \overline{C}_{p}&= \mathbb{E}\left[\log_{2}(1+\gamma)\right]=\int_{0}^{\infty} \log_{2}(1+\gamma)f_{\gamma_{p}}(\gamma)d\gamma\\
   &=\frac{1}{\log(2)}\int_{0}^{\infty} \frac{\overline{F}_{\gamma_{p}}(\gamma)}{1+\gamma} d\gamma,
\end{aligned}
\end{equation}
where $p\in\{L,E\}$ and $\overline{F}_{\gamma_{p}}$ is the complementary of the CDF.\\
The general average secrecy capacity $\overline{C}_{_S}$ can be defined by
\begin{equation}
\overline{C}_{s} = \begin{cases} \overline{C}_{_R}-\overline{C}_{_{E}}, & \mbox{if } \gamma_{_R}> \gamma_{_E} \\ 0, & \mbox{if } \gamma_{_R}< \gamma_{_E} \end{cases}
\end{equation}
where $\overline{C}_{_L}$ is the average capacity of the main link (between $S$ and $R$) and $\overline{C}_{_E}$ is the  average capacity at the eavesdropper $E$.

\subsection{Ergodic Capacity at the Legitimate Receiver}
By referring to Eqs. (5, 6, and 14), the ergodic capacity at the legitimate receiver can be expressed by:
\begin{equation}
\begin{aligned}
   \overline{C}_{R}
   =& \int\limits_{0}^{\infty}\frac{[(s_{_R}-1)\overline{\gamma}_{_L}]^{s_{_R}}T_{_R}^{\mu_{_{R}}}}{\log(2)B(s_{_R},\mu_{_{R}} )}
   \left(\frac{c_{_R}}{c_{_R}+\mu_{_{R}}\kappa_{_I})}\right)^{c_{_R}}\\
   &\times \sum_{i=0}^{\infty} \frac{({c_{_R}})_{i} (s_{_R} + \mu_{_{R}})_{i} (K_{_R}\mu_{_{R}} \kappa_{_I})^{i} }{i!(\mu_{_{R}})_{i} } \\ &\times \frac{\log(1+\gamma)\gamma^{\mu_{_{R}}+i-1}}{K\gamma +[(s_{_R}-1)\overline{\gamma}_{_L}]^{i+s_{_R}+\mu_{_{R}}}}d\gamma.
   \end{aligned}
\end{equation}
To find a closed form of the aforementioned expression, we rewrite the following  expressions as follow \cite{eTransG}
\begin{equation}
\begin{aligned}
    \frac{1}{(T\gamma +\Phi)^{\eta}}=\frac{1}{\Phi^{\eta} \Gamma(\eta)}G_{1,1}^{1,1} 
\left(  \frac{T}{\Phi}
                          \bigg| \begin{matrix} 1-\eta\\
                                       0 \end{matrix}
\right)
   \end{aligned}
\end{equation}
\begin{equation}
\begin{aligned}
    \log(1+\gamma)=G_{2,2}^{1,2} 
\left(  \gamma
                          \bigg| \begin{matrix} 1,\\
                                       1, \end{matrix} \begin{matrix} 1\\
                                       0 \end{matrix}
\right),
   \end{aligned}
\end{equation}
where $G^{m,n}_{p,q} 
\left( \begin{matrix} - | (\cdot,\cdot) \end{matrix}\right)$ is the Meijer G-Function, $\Phi=(s_{_R}-1)\overline{\gamma}_{_L}, \alpha = \mu_{_{R}}+i,$ and $\eta = i+s_{_R}+\mu_{_{R}}.$\\
Then, by substituting Eq. (17) and Eq. (18) in Eq. (16) and by referring to the [\citenum{Mathematica}, 07.34.21.0011.01], we obtain:
\begin{equation}
\begin{aligned}
   \overline{C}_{_R}=& \frac{[(s_{_R}-1)\overline{\gamma}_{_L}]^{s_{_R}}T_{_R}^{\mu_{_{R}}}}{\log(2)B(s_{_R},\mu_{_{R}} )}
   \left(\frac{c_{_R}}{c_{_R}+\mu_{_{R}}\kappa_{_I})}\right)^{c_{_R}}\\
   &\times \sum_{i=0}^{\infty} \frac{({c_{_R}})_{i} (s_{_R} + \mu_{_{R}})_{i} (K_{_R}\mu_{_{R}} \kappa_{_I})^{i} }{i!(\mu_{_{R}})_{i} } \\ &\times \frac{1}{\Phi^{\eta} \Gamma(\eta)}\times G_{3,3}^{3,2} 
\left(  \frac{T_{_R}}{\Phi}
                          \bigg| \begin{matrix} 1-\eta,~ \\
                                       0, \end{matrix} \begin{matrix} -\alpha,~  \\
                                        -\alpha, \end{matrix}\begin{matrix} 1-\alpha\\
                                       -\alpha \end{matrix}
\right).
   \end{aligned}
\end{equation}
\subsection{Ergodic Capacity at the Eavesdropper}
Using Eq. (13) and Eq. (14), we can write the ergodic capacity as follows
\begin{equation}
\begin{aligned}
\overline{C}_{_E}= \frac{\beta_{_J}^{\alpha_{_J}}}{\Gamma(\alpha_{_J})} \sum_{n=0}^{\nu-1}\sum_{q=0}^{n}{n\choose q}\int\limits_{0}^{\infty}\frac{e^{-\beta_{_I} \gamma}(\beta_{_I} \gamma)^n \Gamma(\Omega)}{(\beta_{_I}\gamma+\beta_{_J})^{\Omega}\log(2)(1+\gamma)n!} d\gamma
 \end{aligned}
\end{equation}

To facilitate the integral calculation, we can perform the following transformation into Fox-H function
\begin{equation}
(\beta_{_I}\gamma+\beta_{_J})^{1-\Omega}=\frac{1}{\beta_{_J}^{\Omega}\Gamma(\Omega)}H_{1,1}^{1,1} 
\left(  \frac{\beta_{_I}}{\beta_{_J}}\gamma
                          \bigg| \begin{matrix} (1-\Omega,1)\\
                                       (0,1) \end{matrix}
\right).
\end{equation}

\begin{equation}
\frac{1}{1+\gamma}=H_{1,1}^{1,1} 
\left(  \gamma
                          \bigg| \begin{matrix} (0,1) \\
                                       (0,1) \end{matrix}
\right).~~ ; ~~e^{-\beta_{_I}\gamma}=H^{1,0}_{0,1} 
\left(  \beta_{_I}\gamma
                          \bigg| \begin{matrix} -\\
                                       (0,1)  \end{matrix}
\right).
\end{equation}

Then, we substitute Eq. (21) and Eq. (22) in Eq. (20) and we compute the integral \cite{TripFoxInt}. Hence, we obtain
\begin{equation}
\begin{aligned}
\overline{C}_{_E}&=\frac{1}{\log(2)\Gamma(\alpha{_J})}\sum_{n=0}^{\nu-1}\sum_{q=0}^{n}{n\choose q}\frac{1}{n!\beta_{_I}\beta_{_I}^{q}} \\
&\times H_{0,1;1,1;1,1}^{1,0;1,1;1,1} 
\left(  \frac{1}{\beta_{_I}},\frac{1}{\beta_{_J}}
                          \bigg| \begin{matrix} (-n;1,1)\\
                                       (-;-) \end{matrix}
                          \bigg| \begin{matrix} (0,1)\\
                                       (0,1) \end{matrix}  
                           \bigg| \begin{matrix} (1-\Omega,1)\\
                                       (0,1) \end{matrix}             
\right).
 \end{aligned}
\end{equation}
where  $H_{m_{1},n_{1};m_{2},n_{2};m_{3},n_{3}}^{p_{1},q_{1};p_{2},q_{2};p_{3},q_{3}}(-|(\cdot,\cdot))$ is the bivariate Fox H-function \cite{eBivariate_Relay,eT}.
Therefore, by substituting Eq. (19) and Eq. (23) into Eq. (15), we obtain the average secrecy capacity as given on the top of the next page.
\begin{figure*}
\begin{equation}
\begin{aligned}
   \overline{C}_{s}=\frac{1}{\log(2)}&\left\{ \frac{[(s_{_R}-1)\overline{\gamma}_{_L}]^{s_{_R}}T_{_R}^{\mu_{_{R}}}}{B(s_{_R},\mu_{_{R}} )}
   \left(\frac{c_{_R}}{c_{_R}+\mu_{_{R}}\kappa_{_I})}\right)^{c_{_R}}
    \sum_{i=0}^{\infty} \frac{({c_{_R}})_{i} (s_{_R} + \mu_{_{R}})_{i} (K_{_R}\mu_{_{R}} \kappa_{_I})^{i} }{i!(\mu_{_{R}})_{i}\Phi^{\eta} \Gamma(\eta) } G_{3,3}^{3,2} 
\left(  \frac{T_{_R}}{\Phi}
                          \bigg| \begin{matrix} 1-\eta,\\
                                       0, \end{matrix} \begin{matrix} -\alpha, \\
                                        -\alpha, \end{matrix}\begin{matrix} 1-\alpha\\
                                       -\alpha \end{matrix}
\right)\right.\\
&-\left. \frac{1}{\Gamma(\alpha{_J})}\sum_{n=0}^{\nu-1}\sum_{q=0}^{n}{n\choose q}\frac{1}{n!\beta_{_I}\beta_{_I}^{q}} 
 H_{0,1;1,1;1,1}^{1,0;1,1;1,1} 
\left(  \frac{1}{\beta_{_I}},\frac{1}{\beta_{_J}}
                          \bigg| \begin{matrix} (-n;1,1)\\
                                       (-;-) \end{matrix}
                          \bigg| \begin{matrix} (0,1)\\
                                       (0,1) \end{matrix}  
                           \bigg| \begin{matrix} (1-\Omega,1)\\
                                       (0,1) \end{matrix}             
\right)\right\}
   \end{aligned}
\end{equation}
\sepline
\end{figure*}
 
 \section{Numerical Results and Discussions}
 In this Section, and before investigating the security performance of using the jammer, we are discussing firstly the impact of the shadowing(light shadowing $L_S$ and dense shadowing $D_S$) and blockage (light blockage $L_B$ and dense blockage $D_B$) on the legitimate receiver link by studying the Rician Shadowed as a special case of $\mathcal{D}(\cdot)$. Under the values mentioned in \cite{ShadRician},  we note that for $(L_{B}, L_{S})$, we have an outage probability of 0.02 at SNR = 20 dB. For the same SNR, the outage probability jumps to more than 0.56  in the case of $(L_{B}, D_{S})$.  Moreover, during the scenario when the transmission link is subject to $(D_{B}, L_{S})$, the outage probability is 0.2 at SNR = 14 dB. However, to maintain the same level of outage (0.2), the communication link needs to satisfy higher SNR of at least 28 dB when the shadowing becomes dense ($D_{B}, D_{S}$). Therefore, we deduce that the shadowing effect dominates the blockage. In other words,  our communication system is sensitive to the shadowing effect much more than the blockage density. 
 
  \begin{figure}[H]
\includegraphics[height=65mm, width=0.95\linewidth]{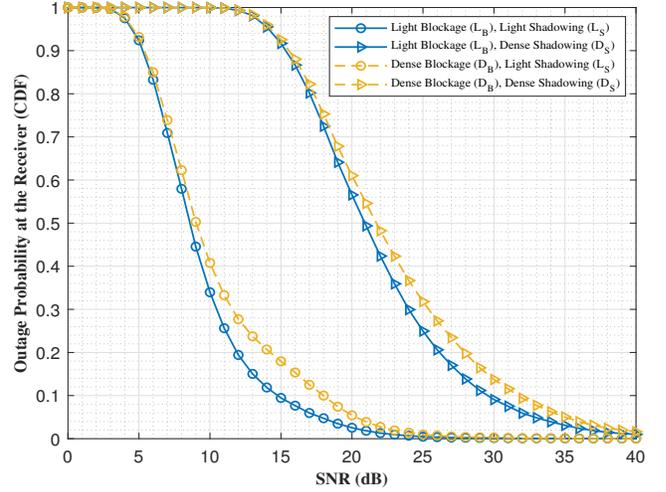}
\caption{Impact of the shadowing and the blockage density on the outage probability at the legitimate receiver. }
\end{figure}

 \begin{figure}[H]
\includegraphics[height=65mm, width=0.95\linewidth]{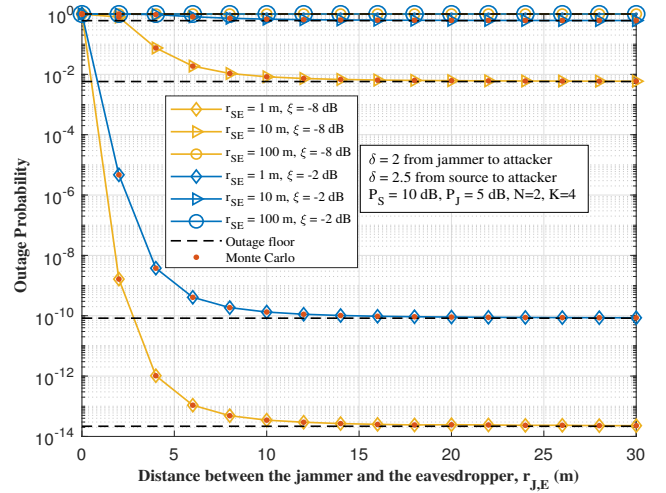}
\caption{Outage probability at the eavesdropper versus distance from the jammer to the eavesdropper $r_{_{J,E}}$.
}
\end{figure}
In Fig. 3, we studied the outage probability at $E$ versus a range of $r_{_{J,E}}\in$ [0 m  30 m], and we fixed   $P_{_J}$ at 5 dB,   $P_{_S}$ = 10 dB. Two different  threshold scenarios were proposed: $\zeta$ = -8 dB and $\zeta$ = -2 dB. If we suppose that to ensure secure communication, the outage probability should be more than $10^{-2}$ for $\zeta$= -8 dB.  In this case,  $r_{_{J,E}}$ should be less than 1 m  when the attacker is very close to the source ($r_{_{S,E}} $= 1 m). If the attacker starts to be 10 m away from $S$, the communication link is protected if $r_{_{J,E}} <$  15 m. In the second scenario, when the security requirements necessitate $\zeta$ = -2 dB, the jammer should be at most 1.5 m away from the eavesdropper to accomplish an outage probability of more than $10^{-2}$ in case that $r_{_{S,E}}$ = 1 m. Both scenarios emphasize on the impact of  the distance separating $J$ from $E$ in the critical case where $r_{_{S,E}} \leq $ 10 m. In those scenarios, the use of the $J$ will be very effective to protect the communication link within a short distance range.  As $r_{_{J,E}}$ increases, the outage cannot decrease anymore and we have an outage floor. This is due to the impact of the path loss on the jamming signal, of course under the aforementioned fading model, power setups, and number of antennas at $J$. Hence, to increase the jamming range of $J$ and mitigate the path loss effect, we have to increase $P_{_S}$ and K, which will be discussed with further details in the Fig. 5.

 \begin{figure}
\includegraphics[height=63mm, width=0.90\linewidth]{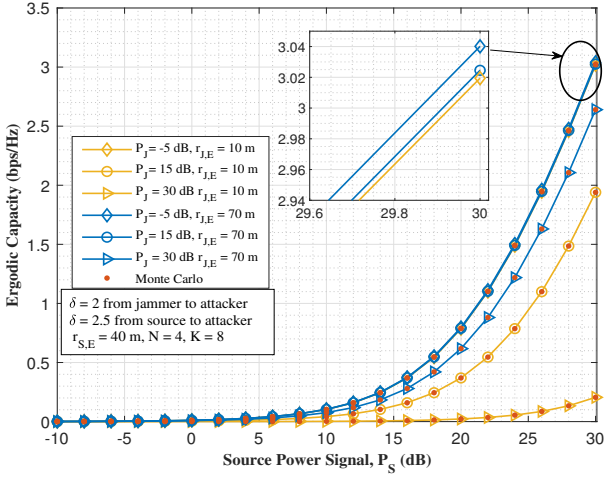}
\caption{Effect of the jamming power signal $P_{J}$ on the ergodic capacity at the eavesdropper end.}
\end{figure}

 Fig. 4 demonstrates the impact of the jamming power generated by the jammer $P_{_J}$ on the ergodic capacity at the eavesdropper $E$ with respect to the source power $P_{_{S}}$. In the first scenario, by fixing  $r_{_{J,E}}$ = 10 m, we can observe that increasing $P_{_J}$ has no effect on the ergodic capacity for  low  transmission  power region $P_{_{S}}\in $[-10 dB, 8 dB], since the average capacity is already null because of the low values of $P_{_{S}}$.  Hence, using a jammer will not make any difference within this power interval. However, for high $P_{_{S}}$  values such as  25 dB and by using $P_{_J}$ = -5 dB,  the ergodic capacity reaches about 1.7 bps/Hz. Therefore, we need to increase the jamming power to at least $P_{_J}$ = 15 dB to decrease the average capacity to 0.95 bps/Hz and to $P_{_J}$ = 30 dB to achieve a rate = 0.05 bps/Hz. In the second scenario, when $r_{_{J,E}}$ = 70 m $> r_{_{S,E}}$ (the attacker is closer to the BS than to the jammer), we remark that for low range of $P_{_{S}}$, the influence of the AN signal is similar to the previous scenario. On the other hand,  for a high  $P_{_{S}}$,  the difference between the rate corresponding to  $P_{_J}$ = -5 dB and $P_{_J}$ = 15 dB is negligible, where we need at least 30 dB at the jammer to decrease the rate from 3.1 bps/Hz to 2.7 bps/Hz. Therefore, the communication could be secured with low jamming power if $r_{_{J,E}} <$ $ r_{_{S,E}}$, otherwise, we need to increase $P_{_{S}}$.
 \begin{figure}
\includegraphics[height=63mm, width=0.9\linewidth]{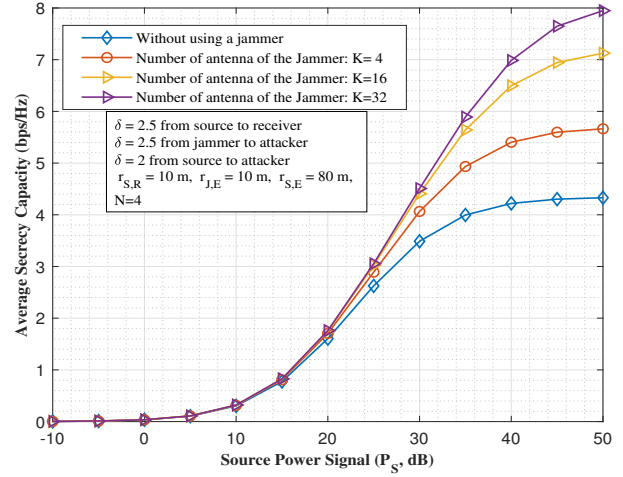}
\caption{Average secrecy capacity for different number of antenna at the jammer.}
\end{figure}

Now, we are focusing of the relation between the study of the average secrecy capacity $\overline{C_{_s}}$ and the number of antennas K, at the jammer.
As shown if Fig. 5 and by fixing the number of antennas at the BS to N = 4, we observe that the average secrecy capacity for different values of K are tightly close within the BS power $\in$ [-10dB  18 dB]. We note that as $P_{_S}$ increases, we start to notice the impact of K on the average secrecy capacity. If we need $\overline{C_{_s}}$ equal to at least 5.5 bps/Hz for $P_{_S}$ = 35 dB, we can only achieve 4 bps/Hz when we are not using any jammer. To satisfy such requirement, we need at least 4 antennas. Therefore, the number of antennas K, has a major impact on the performances of  security scheme.  

 \section{Conclusion}
In this paper, we examined the PLS in the wireless vehicular network while the communication is subject to an eavesdropping attack. We proposed the use of a friendly jammer that will transmit AN to perturb the attacker's channel and decrease its SINR. As a channel model, we adopted the recently proposed  Double $\kappa$-$\mu$ Shadowed Fading model, which provides a variety of fading distribution models.   To evaluate the security performance,   we studied the outage probability and the secrecy capacity with respect to special fading models such as Rician shadowed and Nakagami-m. The results showed that the jammer has a significant impact on the security performance where we can obtain a notable outage probability at the eavesdropping end, especially for short range distance separating the jammer from the attacker.  Moreover, the plots demonstrated the improvement of the average secrecy capacity by equipping the jammer with massive number of antennas.
\section*{Appendix}
To prove that Nakagami-m distribution is a special case of the envelope of  $\mathcal{D}(\cdot)$, we can refer to the envelope equation [\citenum{channel}, Eq. (5)] and substitute   $c_{_l}= s_{_l}= \infty $, $\kappa_{_l}$ = 0 and $\mu_{_l}$ = m. Then we perform the following computation:
\begin{equation}
\begin{aligned}
&\lim \limits_{\substack{c_{_l}\to\infty\\s_{_l}\to\infty}}
{ f}_{X_{_l}}(X) =\lim \limits_{\substack{c_{_l}\to\infty\\s_{_l}\to\infty}}\frac{2(s_{_l}-1)^{s_{_l}}c_{_l}^{c_{_l}}T_{_l}^{\mu_{_l}}X^{2\mu_{_l}-1}{\hat{X}}^{2s_{_l}}}{\left(T_{_l}X^{2}+(s_{_l}-1)
\hat{X}^{2}\right)^{s_{_l}+\mu_{_l}}}\\
&\frac{(c_{_l}+\mu_{_l}\kappa_{_l})^{-c_{_l}}}{B(s_{_l},\mu_{_l})}\times_{2}F_{1}\left( c_{_l}, s_{_l}+\mu_{_l};\mu_{_l}; \frac{K_{_l}\mu_{_l}\kappa_{_l}X}{T_{_l}X+(s_{_l}-1)\hat{X}^{2}}\right)\\
&~~~~=\lim \limits_{s_{_l}\to\infty}
2m^{m}\left( \frac{s_{_l}-1}{\frac{mX^{2}}{\hat{X}^2}+s_{_l}-1}\right)^{s_{_l}}\frac{X^{2m-1}\Gamma(s_{_l}+m)}{\hat{X}^{2(s_{_l}+m)}\Gamma(s_{_l}+m)}\\
&~~~~= \frac{2}{\Gamma(m)}\left(\frac{m}{\hat{X}^{2}}\right)^{m}\exp{\left(-\frac{mX^{2}}{\hat{X}^2}\right)}X^{2m-1}, 
\end{aligned}
\end{equation}

where $X$ is the random variable, $m$ is the shape parameter, $\hat{X}$ is the root mean square  (rms) of the signal envelope,  $T_{_l}$ = m, $K_{_l}$ = 0, $_{2}F_{1}(\cdot,\cdot;\cdot;0)$ = 0 according to the identity: [\citenum{Mathematica}, 07.23.03.0001.01], and $\hat{X}^{2}$ is the  controlling spread parameter.
\bibliographystyle{IEEEtran}
\bibliography{bibliography}
\end{document}